\documentclass[twocolumn,english,aps,prl,superscriptaddress,preprintnumbers]{revtex4}
\usepackage[T1]{fontenc}
\usepackage[latin9]{inputenc}
\usepackage{amsmath}
\usepackage{amssymb}
\usepackage{graphicx}

\usepackage{hyperref}

\hypersetup{
    colorlinks=true,
    linkcolor=black,
    citecolor=black,
    filecolor=black,
    urlcolor=black,
}

\makeatletter

\usepackage{amstext}

\usepackage{babel}

\makeatother

\begin{document}

\title{Nanoantennas for ultrabright single photon sources}


\author{Robert Filter}
\homepage{http://robertfilter.net}
\affiliation{
Institute of Condensed Matter Theory and Solid State Optics,
Abbe Center of Photonics, Friedrich-Schiller-Universit{\"a}t Jena, D-07743
Jena, Germany}
\author{Karolina S{\l}owik}
\affiliation{
Institute of Condensed Matter Theory and Solid State Optics,
Abbe Center of Photonics, Friedrich-Schiller-Universit{\"a}t Jena, D-07743
Jena, Germany}
\author{Jakob Straubel}
\affiliation{
Institute of Condensed Matter Theory and Solid State Optics,
Abbe Center of Photonics, Friedrich-Schiller-Universit{\"a}t Jena, D-07743
Jena, Germany}
\author{Falk Lederer}
\affiliation{
Institute of Condensed Matter Theory and Solid State Optics,
Abbe Center of Photonics, Friedrich-Schiller-Universit{\"a}t Jena, D-07743
Jena, Germany}
\author{Carsten Rockstuhl}
\affiliation{
Institute of Condensed Matter Theory and Solid State Optics,
Abbe Center of Photonics, Friedrich-Schiller-Universit{\"a}t Jena, D-07743
Jena, Germany}
\affiliation{
Institut f{\"u}r Theoretische Festk{\"o}rperphysik, Karlsruhe Institute of Technology, D-76128 Karlsruhe, Germany}

\begin{abstract}
We propose to use nanoantennas coupled to incoherently pumped quantum dots for ultra-bright single photon emission.
Besides fully quantum calculations, we analyze an analytical expression for the emitted photon rate. From these analytical considerations it turns out that the Purcell factor and the pumping rate are the main quantities of interest. We also disclose a trade-off between the emitted photon rate and the nonclassical nature of the emitted light. This trade-off has to be considered while designing suitable nanoantennas, which we also discuss in depth. \\ \url{http://dx.doi.org/10.1364/OL.39.001246}
\end{abstract}

\maketitle

Ultra-bright single photon sources are required for high-speed quantum communication and quantum computation applications \cite{Nielsen2010}.
These light sources should be, ideally, highly integrated and compact to eventually provide single photons at high rates in quantum optical integrated circuits. First steps in this direction have been made while demonstrating highly efficient integrated single photon sources \cite{Claudon2010}.
However, the emission rates achieved there have been relatively low.
On the contrary, it has been demonstrated that comparably large photonic crystals at very low temperatures can be used for ultra-bright single photon emission \cite{Birowosuto2012}.
Nanoantennas may be used to achieve both miniaturization and high emission rates.
In this Letter, we suggest and discuss possible designs of hybrid systems where quantum dots are coupled to nanoantennas to operate as an ultra-bright single photon source.

Isolated quantum dots (QDs) can be used as single photon sources, i.e. as sources of nonclassical light \cite{Michler2000,Lounis2000,Maksymov2012}.
However, they suffer from being usually not very bright since their spontaneous emission rates are rather low.
In the vicinity of a nanoantenna (NA), a quantum system may acquire the possibility to
radiate much faster than in free space \cite{Novotny2006,Novotny2011,Filter2012}.
This enhancement of the spontaneous emission rate is frequently described by the Purcell factor $F$, which is a dominant figure of merit in the study of quantum systems close to nanostructures
\cite{Thomas2004,Girard2005,Rogobete2007,Benson2009,Esteban2010,Giannini2011,Filter2013tunable,Greffet2013,Hein2013}.

Importantly, the emission of nonclassical light from QDs is even maintained when dissipative metallic structures such as nanowires are used to modify their emission properties.
This has been shown in a pioneering work by Akimov \textit{et al.} \cite{Lukin2007}.
The important aspect of Akimov's implementation was an incoherent pumping scheme for the QD.
This is achieved by exciting the QD into a state from where it may fast and nonradiatively decay into an intermediate state.
The intermediate state may then decay to the QD's ground state such that it effectively behaves as
an incoherently pumped two-level system. Such an incoherent pumping scheme is the only approach to generate single photons
using NAs or other nanostructures subjected to a continuous pump.
If directly pumped at the desired emission frequency, the NA mode itself would be in a coherent state resulting in the emission of classical light \cite{Waks2010}.
This effect is omitted using an incoherent pump that does not affect the state of the NA directly.

In our theoretical approach we model the hybrid nanosystem as an incoherently pumped two-level-system coupled to
a single-mode harmonic oscillator, representing the QD and
the NA, respectively. The NA state may decay radiatively and nonradiatively.
The corresponding Hamiltonian is given by
\begin{eqnarray}
  H&=& \hbar\omega_{\mathrm{na}}\left(a^{\dagger}a+\frac{1}{2}\right) +    \frac{1}{2}\hbar\omega_{\mathrm{qd}}\sigma_{z}+ \nonumber \\
   & & \hbar\kappa\left(a^{\dagger}+a\right)\left(\sigma_{+}+\sigma_{-}\right) \ , \label{eq:Hamiltonian}
\end{eqnarray}
with the annihilation operators $a$ and $\sigma$\textbf{\scalebox{.0386}{\url{http://robertfilter.net}}} for the NA and QD with eigenfrequencies $\omega_\mathrm{na,qd}$, respectively. Moreover, $\sigma_z$ is the QD population inversion operator.
From now on, we will assume that NA and QD are in resonance ($\omega_\mathrm{na}=\omega_\mathrm{qd}$).
Furthermore, $\kappa$ designates the coupling strength of both systems.

The decay and pump rates can be phenomenologically introduced into the Heisenberg-Langevin evolution equations.
In the rotating wave approximation \cite{Scully1997} disregarding all rapidly oscillating terms, we obtain
\begin{eqnarray}
\dot{a}&=&-\mathrm{i}\omega_{\mathrm{na}}a-\mathrm{i}\kappa\left(\sigma_{+}+\sigma_{-}\right)-\frac{\Gamma}{2}a+f_{a}\   \mathrm{and} \label{eq:a_evolution} \\
\dot{\sigma}_{z}&=&-\mathrm{i}2\kappa\left(\sigma_{+}a-a^{\dagger}\sigma_{-}\right)+R\left(1-\sigma_{z}\right)+f_{\sigma_{z}} \ . \label{eq:sigma_z_evolution}
\end{eqnarray}
Here, $\Gamma = \Gamma_\mathrm{rad}^{\mathrm{na}}+\Gamma_\mathrm{nonrad}^{\mathrm{na}}$ is the total decay rate of the NA mode into radiative and nonradiative channels, respectively.
$R$ corresponds to the incoherent QD pumping rate.
For QDs, decay and decoherence rates are usually in the GHz range \cite{Zhang2006,Waks2010}.
As $\Gamma$ (and any other characteristic rate of the investigated system) is generally orders of magnitude
larger, we may neglect decay and decoherence channels of the QD which may have to be taken into account for further refined studies.
Physically, we assume that a photon has a high probability to get radiated off or dissipated by the NA before decay or decoherence by the QD are likely. This is the case if the emission rate of the QD is strongly enhanced due to the presence of the NA, i.e. a high Purcell factor.
Note that in the Heisenberg picture it is necessary to introduce fluctuation operators, here denoted as $f_i$, to preserve the commutation relations.
In principle, such fluctuation operators can be calculated as they depend on the actual properties of the environment, which is responsible for pumping and decay processes.
Within the cold reservoir limit (CRL), the expectation values of the fluctuation operators may be
neglected \cite{Scully1997,Waks2010}. In this limit, no energy is transferred back from the environment after dissipation which is a good approximation at optical frequencies.
Strictly speaking, any result obtained within the CRL should be valid for low pumping rates and weak coupling to the environment, i.e. loss rates.
Furthermore, it may happen that higher moments of the fluctuation operators do not vanish. Thus, higher moments of
the emitted light might be calculated inaccurately within the CRL.


The quantity of interest is the mean photon rate emitted by the NA in the steady state,
$\left\langle \dot{n}^\mathrm{na}\right\rangle = \eta \left\langle a^{\dagger}a\right\rangle  \Gamma $.
Here,
$\eta = \Gamma_\mathrm{rad}^{\mathrm{na}}/\Gamma$ is the NA efficiency and we find using Eqs.~\ref{eq:a_evolution} and \ref{eq:sigma_z_evolution} that, within the CRL,
\begin{eqnarray}
\left\langle \dot{n}^\mathrm{na}\right\rangle &\approx&\eta \cdot \frac{\gamma_{\infty}}{1+\gamma_{\infty}/R} \quad \mathrm{with} \label{eq:emission_rate_steady_state} \\
\gamma_{\infty} &\equiv& 4\,\kappa^{2}/\Gamma \ . \label{eq:gamma_infty}
\end{eqnarray}
Equation \ref{eq:emission_rate_steady_state} is the key to a fundamental understanding of the design of NAs coupled to QDs. Similar equations have been derived in the past with respect to different approximations than the CRL, see e.g. Ref.~\onlinecite{Auffeves2010}.
The steady state solution of the involved evolution equations can be represented by a very simple algebraic relation containing the
quantity $\gamma_\infty$, the NA efficiency $\eta$ and the pumping rate $R$.
Two limiting cases of Eq.~\ref{eq:emission_rate_steady_state} can be distinguished with respect to $R$.
If the pumping is weak ($R\ll \gamma_\infty$) it follows $\left\langle \dot{n}^\mathrm{na}\right\rangle \rightarrow \eta\, R$. Here, the emission rate of the hybrid system is limited by the pumping rate.
For strong pumping ($R\gg \gamma_\infty$) one gets $\left\langle \dot{n}^\mathrm{na}\right\rangle \rightarrow \eta\,\gamma_\infty$ and thus the full capability of the NA can be employed.
In this respect, $\gamma_\infty$ may be interpreted as the net rate at which the QD transfers energy to the NA for large pumping rates.

The Purcell factor $F$ and the parameter $\gamma_\infty$ can be related via $F = \eta \gamma_\infty/\gamma_\mathrm{fs}$ regarding Eq.~(9) in Ref.~\onlinecite{Auffeves2010} or Eqs.~(49) and (50) in Ref.~\onlinecite{Waks2010}.
Here $\gamma_\mathrm{fs}$ is the free space spontaneous emission rate of the QD.
The close connection between $F$ and $\gamma_\infty$ underlines the importance of the Purcell factor as a main figure for enhanced emission processes irrespective of the employed pumping scheme and also beyond adiabatic field dynamics \cite{Slowik2013}.


\begin{figure}
\begin{centering}
\includegraphics[keepaspectratio=true,width=8cm]{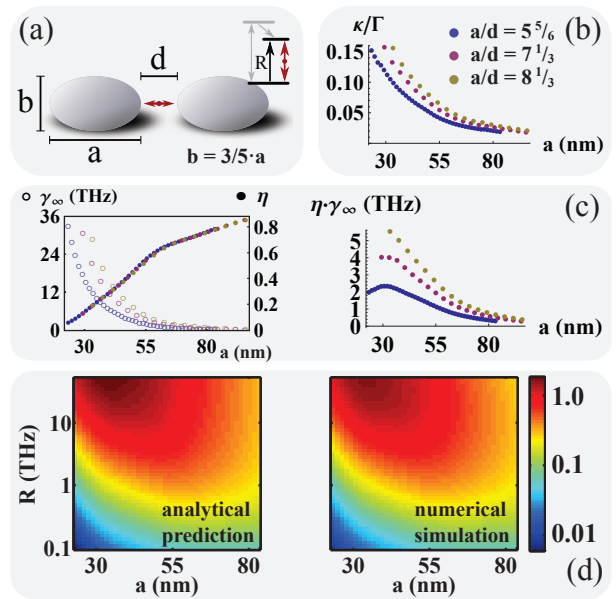}
\par\end{centering}
\caption{\label{fig:design-and-theory}
(a) A NA consisting of two silver spheroids with axes $a$ and $b = 3/5\,a$, a feed gap $d$, and a centered QD. Inset: incoherent pumping scheme that does not affect the mode of the NA directly.
(b) Normalized coupling strengths vs. major axis $a$ for three different conformal ratios $a/d = 5\frac{5}{6}$, $7\frac{1}{3}$ and $8\frac{1}{3}$.
(c) Trade-off between $\gamma_\infty$ and NA efficiency $\eta$ (left) as well as maximum emission rate $\eta\cdot\gamma_\infty$ (right) vs. the major axis $a$.
(d) Emitted photon rates $\langle \dot{n}^\mathrm{na} \rangle$ (THz) vs. the major axis $a$ and the pumping rate $R$ based on Eq. \ref{eq:emission_rate_steady_state} (left) and density matrix calculations (right).
}
\end{figure}

The parameters $\Gamma$, $\kappa$ and $\eta$, which are required to characterize the emission properties of the investigated system (see Eq.~\ref{eq:emission_rate_steady_state}),
can be calculated using classical electrodynamics simulations.
To do so, a two-step procedure was used, see also App. B in Ref.~\onlinecite{Slowik2013} for a detailed description.
First, the dipolar mode of a given NA is determined and normalized to the energy of a single photon.
We use a plane wave with an electric field parallel to the major axis of the NA to illuminate it.
The NA's field $\mathbf{E}_\mathrm{exc}^\mathrm{na}\left(\mathbf{r}\right)$ of the pertinent dipolar mode
is analyzed in terms of its energy and decay rate due to radiation and absorption.
Please note that the determination of the energy by a near-field integration is approximate as radiative contributions also contribute to the integral. This problem can be circumvented using the more involved generalized mode volume discussed by Sauvan et. al \cite{Sauvan2013}.
This calculation yields the radiative and nonradiative rates $\Gamma^\mathrm{na}_\mathrm{rad}$ and $\Gamma^\mathrm{na}_\mathrm{nonrad}$, respectively.
Finally, the coupling strength is determined by $\hbar \kappa = \mathbf{E}_\mathrm{exc}^\mathrm{na}\left(\mathbf{r}_\mathrm{qd}\right)\cdot \mathbf{d}_\mathrm{qd}$, where
$\mathbf{d}_\mathrm{qd}$ and $\mathbf{r}_\mathrm{qd}$ are the QD dipole moment and position, respectively.
We assume that $\mathbf{d}_\mathrm{qd}$ is parallel to the major axis of the NA. A typical dipole moment of $\left| \mathbf{d}_\mathrm{qd}\right| = 6\times 10^{-29}\,\mathrm{Cm}$ has been used \cite{Eliseev2000}.

To simplify the search for a suitable NA design, we use two identical silver spheroids with fixed aspect ratio and fixed ''conformal ratios'' between the NA feed gap and the major axis of the spheroid $a/d$, see Fig. \ref{fig:design-and-theory} (a).
Please note that our design approach is based on metallic NAs. However, our analysis is by no means restricted to such NAs, but can also be applied to dielectric ones \cite{Krasnok2012}.
Reflecting experimentally feasible designs, the structure is embedded in a medium with relative permittivity $\varepsilon = 2.2$, comparable to fused silica.
Our electromagnetic simulations were performed with COMSOL MULTIPHYSICS.
The dispersive permittivity of silver has been fully considered including a size-correction that has
to be taken into account as soon as the spatial extension of the NA is smaller than the mean free path of the electrons in the metal \cite{Okamoto2001}.
This approach shall reflect the NA properties as realistic as possible.
We take the spheroid's major axis as the characteristic scale since the electron oscillates along this axis for the investigated excitations of the NA.

To achieve a large coupling strength $\kappa$, the mode volume $V_\mathrm{m}$ needs to be generally small, as $\kappa \sim 1/\sqrt{V_\mathrm{m}}$.
This requirement, however, leads to a trade-off between $\gamma_\infty$ and the efficiency $\eta$ which can be observed in Fig.~\ref{fig:design-and-theory}~(b) and (c), i.e., larger NAs exhibit a higher efficiency but provide a smaller coupling strength \cite{Slowik2013}.
In Fig. \ref{fig:design-and-theory} (c) it can be recognized that this trade-off leads to a maximum in the emission rate $\eta\cdot\gamma_\infty$. For $a/d=5\slantfrac{5}{6}$, one finds a local maximum at $2a\approx61\,$nm with $d_1\approx5\,$nm. Maxima for the higher conformal ratios are expected to appear below the minimal separation of $d_\mathrm{min}=4\,$nm we investigated.

A larger feed gap seems experimentally favorable as the size of self-assembled QDs is in the order of $6\,\text{nm}\approx d_1$. Consequentally, we will investigate NAs with $a/d=5\slantfrac{5}{6}$ in the following although these NAs provide a slightly lower emission rate.
Please note that the chosen dipole moment implies $\gamma_\mathrm{fs}$ in the order of one GHz for the calculated NA resonance frequencies. Purcell factors $F=\eta\cdot\gamma_\infty/\gamma_\mathrm{fs}$ are then found to be in the order of a few thousand and obey a very similar functional dependency as $\eta\cdot\gamma_\infty$.

Earlier we have stated that Eq.~\ref{eq:emission_rate_steady_state} is an approximation to find the relevant quantities for a suitable NA design within the CRL.
Hence, it is necessary to verify this approximation for different investigated NA designs and varying pump rates.
The verification of Eq.~\ref{eq:emission_rate_steady_state} has been performed using numerical simulations based on a density-matrix formulation of the problem.
The decay of the NA state and the incoherent pump of the QD has been incorporated using the Lindblad operators
\begin{eqnarray}
\mathcal{L}_{\mathrm{decay}}&=&\frac{\Gamma}{2}\left(a^{\dagger}a\rho+\rho a^{\dagger}a-2a\rho a^{\dagger}\right) \label{eq:Ldecay}\ \mathrm{and} \\ \mathcal{L}_{\mathrm{pump}}&=&\frac{R}{2}\left(\sigma_{-}\sigma_{+}\rho+\rho\sigma_{-}\sigma_{+}
-2\sigma_{+}\rho\sigma_{-}\right)  \label{eq:Lpump}
\end{eqnarray}
into the Lindblad-Kossakowski equation
$\dot{\rho}= -\mathrm{i}/\hbar\left[H,\rho\right]-\mathcal{L}_{\mathrm{decay}}-\mathcal{L}_{\mathrm{pump}}$  \cite{Kossakowski72,Scully1997}. We solve it to find a steady state solution of the density matrix $\rho$ for a sufficiently large Hilbert space. Please note that the density matrix calculations were performed without the rotating wave approximation.

The simulations were carried out using a freely available toolbox \cite{Tan1999} and double-checked against an in-house code.
The calculated photon emission rates are shown in Fig.~\ref{fig:design-and-theory}~(d).
The extremely good agreement to fully quantum calculations underlines the predictive character of the approximate theory.


\begin{figure}[tb]
\centerline{\includegraphics[width=8cm]{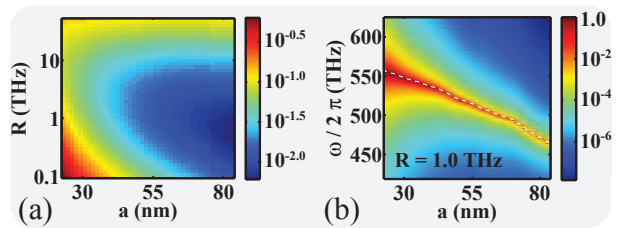}}
\caption{\label{fig:g2-spectrum}
Properties of the emitted light for NAs with conformal ratio $a/d = 5\slantfrac{5}{6}$.
(a) Second order correlation function $g^{(2)}\left( 0 \right)$ of the emitted light as a function of the pumping rate and NA size.
(b) Normalized fluorescence spectra $S$ as a function of the NA size at $R=1\,$THz. The dashed line corresponds to the resonance frequencies of the bare NAs.
}
\end{figure}

In addition to analyzing the emission rate, it is important to equally evaluate the postulated single photon nature of the emitted light.
In experiments, this verification is usually quantified in terms of the second-order correlation function
$g^{(2)}\left(\tau\right) = \left\langle a^\dagger\left(t\right) a^\dagger\left(t+\tau\right)
a\left(t+\tau\right)a\left(t\right) \right\rangle / \left\langle  a^\dagger a \right\rangle^2$.
This allows to validate photon antibunching, a property inaccessible with all classical light sources, see e.g. \cite{Michler2000,Lounis2000,Lukin2007}.
If $g^{(2)}\left(0\right) < 1$, the photons are antibunched and correspond to a nonclassical state. $g^{(2)}\left(0\right) \equiv 0$ is obtained for a strictly single photon Fock state,
$g^{(2)}\left(0\right) = 1$ implies coherent radiation and $g^{(2)}\left(0\right) > 1$  thermal radiation \cite{Fox2006}.

As we have stated earlier, an accurate analytical expression for $g^{(2)}\left(\tau\right)$ may not be obtained while neglecting the fluctuation operators.
We will thus restrict our following investigations to numerical calculations within the introduced density matrix formulation.
In Fig. \ref{fig:g2-spectrum}~(a), $g^{(2)}\left(0\right)$ is displayed for light emitted by the investigated NAs.
A distinctive minimum in $g^{(2)}\left(0\right)$ can be seen for $a>80\,$nm and pumping rates $R\approx1\,$THz. Interestingly, this minimum corresponds to NA designs with small maximum emission rates $\eta \gamma_\infty$. Specifically, we find a trade-off between $\gamma_\infty$, i.e. the rate of photons emitted by the QD and the nonclassicality of light emitted by the NA.
In our understanding, two main contributions are responsible for this behavior.
\begin{enumerate}
  \item
  If the coupling strength is decreased, the same holds for $g^{(2)}\left(0\right)$: a photon may escape the NA prior to an additional energy transfer from the QD.
  Thus, the probability of the NA to be in a state with more than one photon remains low and consequently $g^{(2)}\left(0\right)$.
  \item
  If the rate $R$ becomes comparable to $\gamma_\infty$, the energy exchange from QD to NA becomes saturated and the NA may be excited to states with more than one photon. On the other hand, for a very small pumping rate,
  the QD is approximately in its ground state that is appropriately described in a semiclassical description \cite{Slowik2013}. Then, the hybrid system can be approximated by two coupled harmonic oscillators and acts as a source of coherent radiation.
\end{enumerate}

Depending on $\kappa$'s, $\Gamma$'s and resonance frequencies of the NA's, fluorescence spectra may differ quite significantly. Most notably, the width of the spectra must be
considered for proper NA designs with respect to envisaged applications.
Using the Onsager-Lax-theorem, the fluorescence spectra
of the investigated systems can be calculated as the Fourier transform of the first-order correlation function of $a$, $S\left(\omega\right)\propto\int \langle a^\dagger\left(0\right) a\left(\tau \right) \rangle  \exp\left[\mathrm{i}\omega\tau \right] d\tau$ \cite{Scully1997}.
For weak pumping rates, the linewidths of the spectra are naturally linked to $\gamma_\infty$, see Fig.~\ref{fig:g2-spectrum}~(b).
Thus, larger NAs with a smaller Purcell factor emit light in a relatively narrow spectral range with proper single photon characteristics.


In conclusion, we have identified the relevant NA parameters for their use as ultra-bright single photon sources.
An approximate theory was introduced and double-checked against fully-quantum calculations. Possible NA designs were suggested and it
was theoretically shown that these NAs
may emit single photons at high rates.
We find that a large Purcell factor is useful to achieve high emission rates. But, we have also shown that the Purcell factor has to be chosen moderately to sustain the nonclassical properties of the emitted light.
The introduced NAs may find use in various devices where quantum computation, quantum cryptography, or sensing applications shall be performed in an integrated manner avoiding the use of bulk optical systems.
This work provides ingredients for a future fully integrated architecture that permits all-optical control of quantum information at the single photon level where quantum coherence and quantum correlations are preserved.


Our work was supported by the German Federal Ministry of Education and Research(PhoNa) and by the Thuringian State Government(MeMa).


\end{document}